\documentclass[a4paper, 12pt]{article}
\author{C. J. de Matos\footnote{ESA-HQ, European Space Agency, 8-10 rue Mario Nikis, 75015 Paris, France, e-mail: Clovis.de.Matos@esa.int}
}
\title{Generation of Closed Timelike Curves with Rotating Superconductors}

\begin{document}

\maketitle \begin{abstract} The spacetime metric around a rotating
SuperConductive Ring (SCR) is deduced from the gravitomagnetic
London moment in rotating superconductors. It is shown that
theoretically it is possible to generate Closed Timelike Curves
(CTC) with rotating SCRs. The possibility to use these CTC's to
travel in time as initially idealized by G\"{o}del is
investigated. It is shown however, that from a technology and
experimental point of view these ideas are impossible to implement
in the present context.
\end{abstract}

\section{Introduction}

In Newtonian physics, causality is enforced by a relentless
forward march of an absolute notion of time. In special relativity
things are even more restrictive; not only must you move forward
in time, but the speed of light provides a limit on how swiftly
you may move through space (you must stay within your forward
light cone). In general relativity it remains true that you must
stay within your forward light cone; however this becomes strictly
a local notion, as globally the curvature of spacetime might
"tilt" light cones from one place to another. It becomes possible
in principle for light cones to be sufficiently distorted that an
observer can move on a forward directed path that is everywhere
timelike and yet intersects itself at a point in its "past"-this
is a Closed Timelike Curve (CTC) \cite{Carroll} \cite{Godel1949}.

\section{Closed Timelike Curves in Stationary, Axisymmetric Metrics}

The general metric for a stationary, axisymmetric solution of
Einstein Field Equations (EFE) containing CTCs, with rotation is
given by \cite{Lobo2003} \cite{Tipler1974}:
\begin{equation}
ds^2=-A(r)c^2 dt^2+2B(r)rcd\phi
dt+C(r)r^2d\phi^2+D(r)(dr^2+dz^2)\label{3}
\end{equation}
The range of the coordinates is: $t\in]-\infty, +\infty[$,
$r\in[0, +\infty[$, $\phi\in[0, 2\pi[$, and $z\in]-\infty, +
\infty[$, respectively. The metric components are functions of $r$
alone. The determinant of the metric tensor is assumed to be
Lorentzian,
\begin{equation}
g=det(g_{\mu \nu})=-(AC+B^2)D^2<0,\label{00}
\end{equation}
therefore
\begin{equation}
(AC+B^2)>0.\label{0}
\end{equation}
Landau demonstrated that condition, Equ.(\ref{00}), is always
fulfilled in physically real spacetime \cite{Landau}
\cite{Hawking1991}. If the metric becomes non-Lorentzian,
spacetime becomes unstable \cite{Yahalom} and decays in very short
intervals of time to the Minkowsky metric.

 Since the angular coordinate, $\phi$, is periodic, an azimuthal curve $\gamma
=\{t=Cte, r=Cte, z=Cte\}$ is a closed curve of invariant length
$s^2_\gamma=C(r)(2\pi)^2$. If $C(r)$ is negative then the integral
curve with $(t,r,z)$ fixed is a CTC.

\section{Gravitomagnetic London Moment and Spacetime Metric around a Rotating SCR}
When a SCR is set into rotation it generates a gravitomagnetic
field, call gravitomagnetic London moment \cite{de
Matos2}\cite{Tajmar2006}\cite{TajmarII2006}.
\begin{equation}
B_g=\frac{\rho^*}{\rho} 2 \omega=\frac{\rho^*}{\rho}\nabla\times
v_\phi\label{GMLM}
\end{equation}
Where $\rho^*$ is the Cooper pairs mass density, $\rho$ is the SC
bulk mass density, $\omega$ is SC's angular velocity, and $v_\phi$
is the tangential velocity of the SCR along the azimuthal
direction. For commodity we define the Cooper pair fraction,
$a=\rho^*/\rho$. This London-type gravitomagnetic field is
constant within the interior surface of the ring. Since the
gravitomagnetic field is originated from a vector potential $A_g$
\begin{equation}
B_g=\nabla\times A_g
\end{equation}
We deduce that in a superconductor the gravitomagnetic vector
potential is proportional to the azimuthal velocity of the ring.
\begin{equation}
A_g=a v_\phi=a r \omega\label{1}
\end{equation}
Where $r$ is the radial distance from the SCR's rotation axis.
From the weak field approximation of EFE, which leads to the laws
of gravitoelectromagnetism \cite{Mashhoon}, we know that the
gravitomagnetic vector potential determines the $g_{0i}$
components of the metric tensor.
\begin{equation}
g_{0i}=\frac{4}{c}A_{gi}\label{2}
\end{equation}
Doing Equ.(\ref{2}) into Equ.(\ref{1}) we obtain
\begin{equation}
g_{0\phi}=\frac{4}{c} a r \omega.
\end{equation}
Assuming a SCR with height much larger than its radius, $R$, which
is equivalent to the assumption of an hollow infinitely long
superconductive cylinder, we have no gravitomagnetic field outside
the cylinder. Therefore knowing $g_{0\phi}$, and imposing a flat
metric outside the SCR, i.e. for $r\geq R$, we deduce the other
metric components in Equ.(\ref{3}).
\begin{eqnarray}
A(r)=1 \label{a} \\B(r)=\frac{4}{c} a r \omega \label{b} \\
C(r)=1-8 a \label{c} \\D(r)=1 \label{d}
\end{eqnarray}
The relativistic interval is:
\begin{equation}
ds^2=-c^2 dt^2+\Big(8 a \frac{\omega r}{c} \Big) r c d\phi
dt+\Big(1-8 a \Big)r^2d\phi^2+dr^2+dz^2\label{4}
\end{equation}
which as expected from the boundary conditions defined above,
simplifies to the flat metric in the limit where $r=R$ and
$\omega=d\phi/dt$:
\begin{equation}
ds^2=-c^2 dt^2+R^2d\phi^2+dr^2+dz^2\label{5}
\end{equation}
As we mentioned above, when the azimuthal
metric component becomes negative CTCs become possible in certain
regions.
\begin{eqnarray}
1-8 a<0\\
a>\frac{1}{8}\label{6}
\end{eqnarray}
Therefore when the Cooper pair fraction in the superconductive
material is higher than $1/8$, azimuthal closed curves $\gamma
=\{t=Cte, r=Cte, z=Cte\}$, designated by CTCs, are generated when
the superconductor is set rotating. In general the Cooper pair
fraction in common SCs is $a\sim10^{-7}$. Therefore a Cooper pair
fraction higher that $1/8$ is extremely challenging and is not
achievalble presently with any known superconductor. Making
abstraction of current technological limitations, let us ask in
what region of space, with respect to the SCR's rotation axis,
will the CTCs be located? To answer this question we need to
evaluate the constraints imposed by having a Lorentzian metric
determinant \label{00}.

Doing Equ.(\ref{a})-Equ.(\ref{c}) into Equ.(\ref{0}), we find five
different cases: For $a\leq1/8$, the SCR cannot generate CTCs and
the metric Equ.(\ref{4}) will be allowed for all $r>0$. In the
case, in which the SCR is capable to generate CTCs, i.e., $a>1/8$
we have four possibilities depending on the value of $a$ and of
the angular velocity, $\omega$:
\begin{enumerate}
\item  \label{approx_1} If $1/8<a<1$ and $\omega<1/a$ leads to $r>r_{max}$.

\item  \label{approx_2} If $1/8<a<1$ and $\omega>1/a$ leads to $r<r_{max}$.

\item  \label{approx_3} If $a>1$ and $\omega>1/a$, then $r>r_{max}$.

\item  \label{approx_4} If $a>1$ and $\omega<1/a$, then
$r<r_{max}$.
\end{enumerate}
Where
\begin{equation}
r_{max}=\frac{c}{4a\omega}(8a-1)^{1/2}\label{max}
\end{equation}
From Equ.(\ref{max}) we see that for conditions close to the
boundary conditions of case \ref{approx_1}., \ref{approx_2}.
\ref{approx_3}., and \ref{approx_4}., the radius, $r_{max}$, is
approximately equal to the distance between the Earth and the
Moon. Therefore in the case \ref{approx_1}., and \ref{approx_3},
the metric Equ.(\ref{4}) is not Lorentzian inside any SCR having a
realistic radius ($R<<r_{max}$). In these cases the metric
Equ.(\ref{4}) decays to Minkowsky metric and $B_g=0$ for
$r<r_{max}$. For the case \ref{approx_2}., and \ref{approx_4}.,
however, SCR's with realistic size would be capable to host CTC's
in their hollow region (if the challenging Cooper pair fraction
could be achieved). The light cone structure and the requirements
to use these CTCs for traveling in time will be investigated in
the next two sections.

\section{Lightcone Structure Along the Azimuthal Direction}
In the examination of the lightcone structure, we will see in what
follows that the azimuthal closed $\phi$-curves (note that since
we are here interested in lightcones $dt\neq 0$) are indeed
spacelike for certain values of $a$ and timelike for others. Doing
$dr=dz=0$ in Equ.(\ref{4}), for the case of lightcones, $ds=0$
\begin{equation}
-c^2 dt^2+\Big(\frac{8a \omega r}{c} \Big) r c d\phi
dt+\Big(1-8a\Big)r^2d\phi^2=0\label{7}
\end{equation}
Solving Equ.(\ref{7}) with respect to the variable $cdt/r d\phi$
we obtain.
\begin{equation}
\frac{cdt}{r d\phi}=-\frac{1}{2}\Bigg(-\frac{8a\omega r}{c}\pm
\sqrt{\Big(\frac{8a\omega
r}{c}\Big)^2+4\Big(1-8a\Big)}\Bigg)\label{8}
\end{equation}

For $0 \leq a<\frac{1}{8}$, which also includes the case of a
non-superconductive material, $a=0$, we have
\begin{equation}
\frac{cdt}{rd\phi}\sim \pm1\label{9}
\end{equation}
the lightcone is just the usual Minkowskian one.

For $a=1/8$, we have:
\begin{equation}
\frac{cdt}{rd\phi}= \left \{\begin{array}{ll}
0\\
\omega r/c
\end{array} \right.
\end{equation}

the lightcone becomes very narrow, since in general $\omega r<<c$,
it also dips and touches the $\phi$ axis.

For $a>1/8$; the $\phi$-curve is enclosed within the lightcone:
\begin{equation}
\frac{cdt}{rd\phi}=\frac{8 a \omega r}{2c}\pm \epsilon\label{11}
\end{equation}
Where $\epsilon > \frac{8 a \omega r}{2c}$. The lightcone is still
very narrow. To enlarge it we would need to have $r\omega\sim c$,
which also requires $a>>1/8$ and $v_{\phi}\simeq c$. The curve is
always timelike, and hence the propertime flows monotonically and
never becomes imaginary, i.e. the curve does not reverse and
proceed into the past lightcone. This timelike curve returns to
its original location in spacetime, it is a closed timelike curve,
as we expected for the considered Cooper pair fraction.

In summary when the Cooper pair fraction is important enough for
the rotating superconductor to generate CTCs, $a>1/8$, then in the
regions indicated in condition \ref{approx_1}., \ref{approx_2}.,
\ref{approx_3}., and \ref{approx_4}., the light cone will
\emph{open} in the azimuthal direction and will contain time-like
directions for decreasing $t$ pointing into the past, making
travel into the past possible. These time-like circles $\gamma$
are not geodesics. Since the total acceleration of the curve does
not vanish, as we will see in the next section.

\section{Acceleration Requirements for G\"{o}del's time travel}
If we consider the circle $\gamma$ given by
\begin{equation}
x^0=ct=Cte, x^1=r=Cte, x^3=z=Cte\label{13}
\end{equation}
In this case the interval Equ.(\ref{4}) will reduce to:
\begin{equation}
ds^2=\Big(1-8a\Big)r^2d\phi^2=g_{\phi\phi}d\phi^2\label{17}
\end{equation}
with $r$ belonging to the CTC's allowed domains defined in
conditions \ref{approx_1}., \ref{approx_2}. \ref{approx_3}., and
\ref{approx_4}..The first question to ask is wether there exist
any closed time-like geodesics in the spacetime described by the
metric inside the SCR's hole, Equ.(\ref{4}). If so, it would be
possible to execute time travel in a state of free fall. It turns
out that the answer is no (at least for the stationary case, in
which no angular acceleration is communicated to the SCR). We see
that the tangent vector to the circle $\gamma$,
\begin{equation}
\hat{e}_\phi=\frac{\partial}{\partial
x^2}=\frac{\partial}{\partial \phi},\label{14}
\end{equation}
has the length squared
\begin{equation}
\Big(\frac{\partial}{\partial\phi}\Big)^2=\hat{e}_\phi.\hat{e}_\phi=g_{\phi\phi}=g_{22}=(1-8a)r^2\label{20}
\end{equation}
The quadri-acceleration vector $A^j$ is given by
\begin{equation}
A^j=c^2u^{j}_{;k}u^k\label{24}
\end{equation}
For the time-like unit vector
\begin{equation}
u^ju_j=1,\qquad u^j=-\delta^j_2|g_{22}|^{-1/2}\label{21}
\end{equation}
The semicolon in Equ.(\ref{24}) indicates covariant
differentiation. We then obtain that the acceleration $A$ defined
by
\begin{equation}
A^2=g_{jk}A^jA^k\label{22}
\end{equation}
becomes \cite{Ozsvath2003}
\begin{equation}
A=\frac{1}{4}c^2\frac{d\ln|g_{\phi\phi}|}{dr}.\label{23}
\end{equation}
Or doing Equ.(\ref{20}) into Equ.(\ref{23})
\begin{equation}
A=\frac{2 c^2}{r}\label{15}
\end{equation}
We see that $A$ does not vanish (it becomes null as
$r\rightarrow\infty$).

The total integrated acceleration over $\gamma$ is
\begin{equation}
TA(\gamma)=\oint_{\gamma} Ad\tau\label{16}
\end{equation}
Where $A$ is the acceleration at any point of the curve,
Equ.(\ref{15}), and $\tau$ is the elapsed proper time along
$\gamma$,
$d\tau=ds/c=\Big(\Big|g_{\phi\phi}\Big|\Big)^{1/2}d\phi/c$. Notice
that $TA(\gamma)=0$ if and only if $\gamma$ is a geodesic (closed
gravitational field lines like the ones obtained through
gravitomagnetic induction, $\nabla\times\vec{g}=-\dot{\vec{B_g}}$,
are geodesics). Doing Equ.(\ref{17}) and Equ.(\ref{15}) into
Equ.(\ref{16}) we obtain:
\begin{equation}
TA(\gamma)=4 \pi c \sqrt{|1-8a|}\label{18}
\end{equation}
the total acceleration of the $\gamma$ curve, is a measure of the
total variation of velocity $\Delta v$ needed to achieve a
complete G\"{o}del's round trip.

The total elapsed proper time $PT(\gamma$) along $\gamma$ will be:
\begin{equation}
PT(\gamma)=\oint_{\gamma}d\tau=\frac{2\pi r
\sqrt{|1-8a|}}{c}\label{19}
\end{equation}

Therefore a CTC taken as the world line of an observer would
enable her/him to travel into her/him own past if the acceleration
were tolerable and the proper time for the round trip was less
than its
lifetime\cite{Malament1984-1}\cite{Malament1984-2}\cite{Malament1987}.

\section{Conclusion}

It seems that the rotation of a superconductive ring is capable to
generate CTCs if the Cooper pair fraction, $a=\rho^* / \rho$,
could be raised above the critical value of $a=1/8$. This is a
huge value with respect to the current Cooper pair fraction of
presently known superconductors, $a\sim10^{-7}$. Even if it
becomes possible to design superconductors having this threshold
of Cooper pairs fraction, we see from conditions \ref{approx_1}.,
and \ref{approx_2}., \ref{approx_3}., and \ref{approx_4}., that
the possible values of Cooper pairs fraction $a>1/8$, and the
SCR's angular velocities $\omega$ are severely constrained by the
fact that the metric must remain Lorentzian. CTCs will only be
present inside the SCR's hole, with radius $R<<r_{max}$, for
values of $(a, \omega)$ defined by conditions \ref{approx_2}., and
\ref{approx_4}. In other words conditions \ref{approx_1}., and
\ref{approx_3}, form non-Lorentzian barriers which constrain the
operation of the SCR in terms of allowed Cooper pair fraction and
angular velocities. Outside this allowed regime of operation,
spacetime would become unstable, and would decay to the
Minkowskian metric in a very short time \cite{Yahalom}. This would
also mean that the gravitomagnetic London moment,
Equ.(\ref{GMLM}), would be null in the non-Lorentzian regions.

The utilization of these CTC's to travel in time as G\"{o}del
first idealize seems to be unpracticable, since the total
accelerations needed to perform the time travel, Equ.(\ref{15}),
is not technically achievable, and the total elapsed time for a
round trip is negligibly small, Equ.(\ref{19}).

\end{document}